\documentclass[%
 reprint,
 showpacs,preprintnumbers,
 amsmath,amssymb,
 aps,
 prb,
]{revtex4-1}
\usepackage{dcolumn}
\usepackage{subfigure}
\usepackage{here}
\usepackage{graphicx,color}
\usepackage{mathrsfs}

\makeatletter
\def\btt#1{\texttt{\@backslashchar#1}}
\DeclareRobustCommand\bblash{\btt{\@backslashchar}} \makeatother

\begin{document}

\title{Pseudo-spin triplet superconductivity in transition-metal dichalcogenide monolayers and Andreev reflection in the lateral heterostructures of 2$H$-NbSe$_2$}

\author{Tetsuro Habe}
\affiliation{Department of Applied Physics, Hokkaido University, Spporo, Hokkaido 060-0808, Japan}

\date{\today}

\begin{abstract}
We study the pseudo-spin of electron pair in superconducting transition-metal dichalcogenide monolayers and show that the pseudo-spin affects the electric transport property of lateral heterojunction of the superconducting and metallic monolayers.
The pseudo-spins of two electrons forming a Cooper pair are parallel to each other unlike the real spin being anti-parallel.
In the lateral heterojunction, the electronic transport with forming a Cooper pair, the Andreev reflection, is suppressed with the Fermi level crossing the valence band near the edge in the metallic monolayer.
We numerically investigate the electric transport property of the lateral heterojunctions of semiconducting and superconducting transition-metal dichalcogenides, MoSe$_2$ and NbSe$_2$ monolayers with the charge doping, respectively.
We find the sign change of conductance difference between the normal and superconducting phases by varying the charge density and show that the sign change is resulted from the pseudo-spin triplet superconductivity.
\end{abstract}

\maketitle
\section{Introduction}
Transition-metal dichalcogenides (TMDCs) are atomic layered materials composed of transition-metal and chalcogenide atoms.
The monolayer crystal can be fabricated experimentally by using chemical vapor decomposition (CVD) or cleaving from a single crystal\cite{Helveg2000,Mak2010,Coleman2011,Lee2012,Dong2017}.
The TMDC monolayers show several phases of condensed matter; the superconductivity,\cite{Lu2015,Xi2015,Wang2017}, the charge-density wave,\cite{Ugeda2015,Xi2015} and the topological insulator\cite{Tang2017}.
NbSe$_2$ is also a transition-metal dichalcogenides (TMDC) well known as a conventional superconductor.\cite{Kershaw1967}
The superconductivity has been observed even in the monolayer crystal.\cite{Wang2017}
In the superconducting monolayer, two electrons forming the Cooper pair have opposite spins like conventional superconductors but the spin axis is locked in the out-of-plane direction due to the spin-orbit coupling unlike those.\cite{Xi2015}
This spin character of pair provides unique property to the superconductivity, e.g., the large and anisotropic upper critical field exceeding the Pauli limit.\cite{Xi2015,Wang2017}
Moreover, the other spin-related phenomena in the superconducting NbSe$_2$ monolayer have been studied in several works.\cite{David2018,Rahimi2018,Sohn2018,Shaffer2019,Glodzik2019}

TMDC monolayers have the other internal degree of freedom so-called pseudo-spin which is defined for representing electron states and used to describe the Berry curvature related phenomena, e.g., the valley, spin, and anomalous Hall effects.\cite{Xiao2012,Shan2013,Zibouche2014,Habe2017}
The pseudo-spin represents two Wannier orbitals as two spin elements.
In two Fermi pockets around the K and K$'$ points, the electronic states are approximated to be those of Dirac fermion in the pseudo-spin space.
Then the pseudo-spin varies with the magnitude and direction of the wave vector in the two pockets.\cite{Xiao2012,Cappelluti2013}
Thus the Cooper pair formed of these electrons has the pseudo-spin in the superconducting state.

\begin{figure}[htbp]
\begin{center}
 \includegraphics[width=70mm]{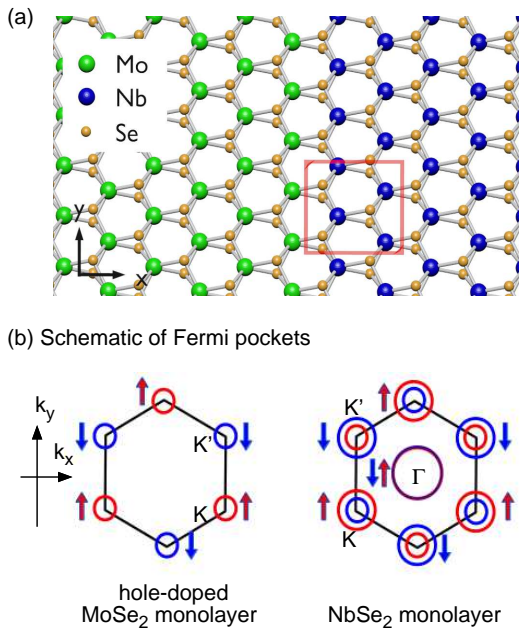}
\caption{
The schematics of (a) lateral heterojunction and (b) Fermi pockets in each TMDC monolayer.
The $y$-axis is taken to be parallel to the interface in (a). The red box indicates the extended unite cell for the numerical calculation in Sec.\ \ref{Sec_numerical}.
In (b), the circles represent the Fermi pocket in Brillouin zone, the hexagonal flame, and the arrows express the spin direction of electrons in each Fermi pocket. 
 }\label{fig_schematic}
\end{center}
\end{figure}

In this paper, we discuss the pseudo-spin of electrons forming Cooper pairs in the superconductor of TMDC monolayer and the effect to the electronic transport property in the lateral heterojunction of the superconducting and metallic TMDC monolayers in Fig. \ref{fig_schematic} (a).
In Sec.\ \ref{Sec_analyze_junction}, we consider an effective model describing electronic states around the K and K$'$ valleys and analyze the pseudo-spin of electrons forming Cooper pairs in the superconducting phase.
Moreover, we calculate the transmission and reflection coefficients in the lateral heterostructure and discuss the effect of pseudo-spin to these coefficients.
In Sec.\ \ref{Sec_numerical}, we calculate the electric conductance, which is associated with the coefficients, in the lateral heterojunction of MoSe$_2$ and NbSe$_2$ monolayers by using a first-principles band calculation and lattice Green's function method.
The discussion and the conclusion are given in Sec.\ \ref{Sec_discussion} and\ \ref{Sec_conclusion}, respectively.

\section{Effective model}\label{Sec_analyze_junction}
We investigate the pseudo-spin texture of the superconducting state in TMDC monolayers.\cite{Lu2015,Xi2015,Wang2017}
The Cooper pair is a spin singlet and formed of two electrons which are the time-reversal partners.
In NbSe$_2$, the Fermi level is crossing the hole band, and it arranges three Fermi pockets enclosing the $\Gamma$, K and K$'$ points in the Brillouin zone as shown in Fig.\ \ref{fig_schematic} (b), so-called the valley degree of freedom.\cite{Habe2019-1}
In what follows, we consider the pair of electrons in the K and K$'$ valleys because the electronic states have non-trivial pseudo-spin texture in these valleys.

\subsection{Electronic states in a TMDC monolayer}
To analyze electronic states at the two valleys, we consider an effective model describing Dirac fermions.
The effective model is represented by a $2\times2$ Hamiltonian,
\begin{align}
H_{\mathrm{eff}}=
\begin{pmatrix}
m+\varepsilon_F&v(\tau k_x-i k_y)\\
v(\tau k_x+i k_y)&-m-s\tau b_0+\varepsilon_F
\end{pmatrix},\label{eq_massive_Dirac}
\end{align}
defined on the basis of two $d$-orbitals, $(|d_{3z^2-r^2}\rangle,\; |d_{xy}\rangle+i\tau|d_{x^2-y^2}\rangle)$, in transition-metal atoms.\cite{Xiao2012}
Here the in-plane wave number $(k_x,k_y)$ is defined with respect to the valley center, and $\tau$ is the valley index which is 1 in the K valley and -1 in the K$'$ valley.
The spin-orbit coupling (SOC) is represented by the Zeeman-like term, $s\tau b_0$, with the coupling constant $b_0$ and the spin index $s=\pm1$ in the $z$ direction.
The other parameters $m$, $\varepsilon_F>0$, and $v$ are the gap energy, the Fermi energy, and the velocity, respectively.
The Hamiltonian can be represented by a superposition of the $2\times2$ identity matrix and the Pauli matrix $\sigma_\mu$ for $\mu=x$, $y$, and $z$ which can be considered as a pseudo-spin operator.
When the Hamiltonian is given by $\boldsymbol{d}\cdot\boldsymbol{\sigma}$ without the identity matrix component, we represent the axis of pseudo-spin by $\boldsymbol{d}/|\boldsymbol{d}|$.

The electronic states consists of a plane wave component and a vector component expressing the pseudo-spin direction.
We consider the electronic states at the Fermi level because the Cooper pair is consisting of two electrons near the level.
The pseudo-spin component is represented by
\begin{align}
|\psi(\varepsilon_F,\theta_k)\rangle=
\frac{1}{\sqrt{1+r^2}}
\begin{pmatrix}
-{\tau}re^{-i\tau\theta_k}\\
1
\end{pmatrix},
\end{align}
with $r=vk_F/(\varepsilon_F+m)$, where $\theta_k$ is defined by $\theta_k=\mathrm{atan}(k_y/k_x)$ and $k_F$ is the Fermi wave number obtained from
\begin{align}
\mathrm{det}[H_{\mathrm{eff}}]=(\varepsilon_F+m)(\varepsilon_F-m-s\tau b_0)-v^2k_F^2=0.
\end{align}
When the electron forms a Cooper pair, the other electron is the time-reversal partner, i.e., it occupies the electronic state transformed by time-reversal operation.
The partner has the same pseudo-spin component because Eq.\ (\ref{eq_massive_Dirac}) is unchanged under time-reversal operation $\mathcal{T}$ with $\mathcal{T}H(s,\tau,\boldsymbol{k})\mathcal{T}^\dagger=H^\ast(-s,-\tau,-\boldsymbol{k})$.
Then the two electrons of Cooper pair have the same pseudo-spin, i.e., the electrons form a pseudo-spin triplet pair, although they have the opposite real spin.

The Cooper pair is always formed of equal pseudo-spin electrons but the polarizing direction changes with the Fermi energy.
We consider two limits of Fermi energy; $m/vk_F\rightarrow\infty$ and $m/vk_F\rightarrow0$, i.e., $r\rightarrow0$ and $r\rightarrow1$, respectively.
The former limit implies that the Fermi level is close to the valence band edge.
In this case, the pseudo-spin is nearly independent of the wave number and has a unique polarizing axis indicated by the pseudo spin component,
\begin{align}
|\psi(\varepsilon_F,\theta_k)\rangle\simeq\begin{pmatrix}
0\\
1
\end{pmatrix}.
\end{align}
In the latter case, the pseudo-spin varies with the wave number and has a helical texture as expressed by the vector component,
\begin{align}
|\psi(\varepsilon_F,\theta_k)\rangle\simeq\frac{1}{\sqrt{2}}\begin{pmatrix}
-\tau e^{-i\tau\theta_k}\\
1
\end{pmatrix}.
\end{align}
\begin{figure}[htbp]
\begin{center}
 \includegraphics[width=80mm]{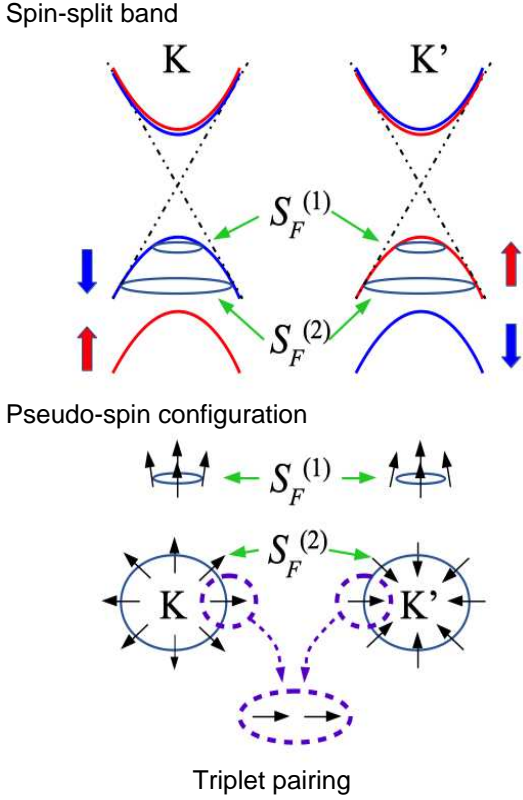}
\caption{
The schematics of spin-split band and its pseudo-spin texture in the Fermi surface $S_F^{(j)}$.
Here, $S_F^{(1)}$ and $S_F^{(2)}$ are the Fermi surface for the Fermi level near and far from the band edge, respectively.
The pseudo-spin in $S_F^{(1)}$ is polarized in the $z$ direction and that in $S_F^{(2)}$ forms a hedgehog in the $xy$ plane.
The dashed lines explain the formulation of triplet Cooper pair.
 }\label{fig_pseudo-spin}
\end{center}
\end{figure}
The former texture and the latter texture are schematically shown as $S_F^{(1)}$ and $S^{(2)}_F$, respectively, in Fig.\ \ref{fig_pseudo-spin}.
In the figure, the arrows indicate the spin and pseudo-spin directions in the upper and lower panels, respectively.
The direction of pseudo-spin $\boldsymbol{d}$ is projected in the wave number space where $\boldsymbol{d}\cdot\boldsymbol{\sigma}$ gives the eigenvalue of pseudo spin for each electronic state.

\subsection{Lateral heterojunction of superconducting and metallic TMDC monolayers}
To discuss the effect of pseudo-spin texture, we consider the scattering problem in a lateral heterojunction of the superconducting and metallic TMDC monolayers. 
The lateral heterojunction is composed of two TMDC monolayers atomically bonded as a monolayer.\cite{Huang2014,Gong2014,Duan2014,Chen2015,Chen2015-2,Zhang2015,He2016}
When the $y$-axis is taken to be the axis parallel to the interface of heterojunction, $k_y$ is preserved throughout the scattering process.
Thus the conducting state can be represented by an eigen-function of $k_y$, and it is a zero-energy state with respect to the Fermi level.
In this section, we consider the heterojunction with zig-zag interface as shown in Fig.\ \ref{fig_schematic} (a), and thus we can analyze the K and K$'$ valleys separately.

We calculate the wave function in the junction by connecting the wave functions in the two monolayers, and thus we consider the pristine monolayers firstly.
We assume that the Fermi level is crossing the lower band of Eq.\ (\ref{eq_massive_Dirac}) in both of the monolayers and that it is far away from the band edge in the superconducting monolayer.
Since then $m\ll vk$ is fulfilled around the Fermi level in the superconducting monolayer, the effective Hamiltonian can be approximated by that of Weyl fermion,
\begin{align}
H_{\mathrm{Weyl}}(\boldsymbol{k})=\begin{pmatrix}
\varepsilon_F&v(\tau k_x-i k_y)\\
v(\tau k_x+i k_y)&\varepsilon_F
\end{pmatrix},\label{eq_Weyl_Hamiltonian}
\end{align}
where we focus on one of spin state and put the Zeeman-like SOC in $\varepsilon_F$.
At each $k_y$, two electronic states are present if $k_x=\pm p_x$ is real with 
\begin{align}
p_x=\frac{1}{v}\sqrt{\varepsilon_F-v^2k_y^2},
\end{align}
and they can be represented by
\begin{align}
|\varphi^\pm_e(\varepsilon_F,k_y)\rangle=\frac{1}{\sqrt{2}}\begin{pmatrix}
\pm\tau e^{\pm i\tau\theta_p}\\
1
\end{pmatrix},\label{eq_electron_state}
\end{align}
with $\theta_p=\mathrm{atan}(k_y/p_x)$, where the superscript indicates the sign of velocity in the $x$-direction and it is opposite to that of $k_x$.
The positive velocity indicates the right going state in the right side of junction in Fig.\ \ref{fig_schematic} (a).

In the Bogoliubov-de Gennes (BdG) formalism, the quasi-particle state in superconductors is represented by the superposition of the electronic state and the hole state of time-reversal partner. 
The hole state is in the opposite valley and described by $-\mathcal{T}H_{\mathrm{Weyl}}\mathcal{T}^\dagger$, where the Zeeman-like SOC is unchanged due to $s\tau\rightarrow s\tau$ under time-reversal operation $\mathcal{T}$.
The pseudo-spin component is the same as the electron state in Eq.\ (\ref{eq_electron_state}), but the direction of velocity is inverted,
\begin{align}
|\varphi^\pm_h(\varepsilon_F,k_y)\rangle=|\varphi^\mp_e(\varepsilon_F,k_y)\rangle,
\end{align}
where the superscript also indicates the direction of velocity.
The quasi-particle states in the superconducting region are the superposition of these states and described by the BdG Hamiltonian,
\begin{align}
H_{\mathrm{BdG}}=\begin{pmatrix}
\varepsilon_F-vk&\Delta\\
\Delta&-(\varepsilon_F-vk)
\end{pmatrix},\label{eq_eff_BdG}
\end{align}
where the upper and lower components in the basis are the zero-energy electron state and the hole state of the time-reversal partner, respectively.
Here $\Delta=g\langle c_{\uparrow,\boldsymbol{k}}c_{\downarrow,-\boldsymbol{k}}\rangle$ is the order parameter of superconducting state where $g$ is the attractive coupling constant and $\langle c_{\uparrow,\boldsymbol{k}}c_{\downarrow,-\boldsymbol{k}}\rangle$ indicates the Cooper pair amplitude with annihilation operator $c_{s,\boldsymbol{k}}$ of electron in the zero energy state.

The zero-energy quasi-particle state is in the superconducting gap and thus its wave function is an evanescent wave which has a spatial dependence of $\exp[i\kappa x]$ with $\mathrm{Im}[\kappa]>0$. 
The damping ratio $\kappa$ can be obtained from $\mathrm{det}(H_{\mathrm{BdG}})=\left(\varepsilon_F-v\sqrt{\kappa^2+k_y^2}\right)^2+\Delta^2=0$,
\begin{align}
\kappa=\pm\frac{1}{v}\sqrt{\varepsilon_F^2-v^2k_y^2-{\Delta}^2\mp2 i\varepsilon_F{\Delta}}.\label{eq_damping_ratio}
\end{align}
As discussed later, $k_y$ of conducting channel is much smaller than $p_x$ in this problem.
Thus the damping ratio can be approximated by
\begin{align}
\kappa\simeq \pm p_x+i\frac{{\Delta}}{v},\label{eq_damp}
\end{align}
where we choose the decay waves in the positive $x$-direction.
We also approximate the pseudo-spin component of the electron and hole states by those for $\Delta=0$ in Eq.\ (\ref{eq_electron_state}) due to $\Delta/v\ll p_x$.
The vector component of the decay wave function is given by
\begin{align}
|\phi^\pm(\varepsilon_F,k_y)\rangle=|\varphi_e^\pm(\varepsilon_F,k_y)\rangle\mp i|\varphi_h^\pm(\varepsilon_F,k_y)\rangle),
\end{align}
where the first and second terms are corresponding to the upper and lower elements in Eq.\ (\ref{eq_eff_BdG}), respectively.
There are other evanescent waves consisting of only electron or hole wave function $|\phi^{(0)}_j\rangle$ which are originated from the sates in the conduction band by solving Eq.\ (\ref{eq_eff_BdG}) with $vk\rightarrow-vk$.
These evanescent waves are not responsible for the Andreev reflection process and orthogonal to other evanescent waves.
Thus they can be eliminated in the calculation of scattering coefficients as discussed later.

In the metallic monolayer, the Fermi energy is assumed to be crossing the lower band near the band edge.
Thus we use the original form of Hamiltonian $H_{\mathrm{eff}}$ in Eq.\ (\ref{eq_massive_Dirac}) and obtain $k_x=\pm q_x$ by solving $\mathrm{det}[H_{\mathrm{eff}}]=0$,
\begin{align}
q_x=\frac{1}{v}\sqrt{(\varepsilon_F+m)(\varepsilon_F-m-s\tau b_0)-v^2k_y^2}.
\end{align}
The pseudo-spin component of electron state is represented by
\begin{align}
|\psi_e^\pm(\varepsilon_F,k_y)\rangle=
\frac{1}{\sqrt{1+r^2}}
\begin{pmatrix}
\pm\tau re^{\pm i\tau\theta_q}\\
1
\end{pmatrix},
\end{align}
with $\theta_q=k_y/q_x$, where the sign $\pm$ indicates the direction of velocity in the $x$ direction.
Then the pseudo-spin of hole state is given by $|\psi_h^\pm(\varepsilon_F,k_y)\rangle=|\psi_e^\mp(\varepsilon_F,k_y)\rangle$.

The electronic transmission process can be described by the stationary state of scattering wave in the junction.
In the metal side, there are one incident electron wave and two reflected waves in which one is the electron wave and the other is the hole wave.
The amplitudes of reflected electron and hole are the normal and Andreev reflection coefficients, $r^{ee}$ and $r^{he}$, respectively.
Then these wave functions satisfy the boundary condition,
\begin{align}
&|\psi_e^+(\varepsilon_F,k_y)\rangle+r^{ee}|\psi_e^{-}(\varepsilon_F,k_y)\rangle+r^{he}|\psi^{-}_h(\varepsilon_F,k_y)\rangle\nonumber\\
=&c_+|\phi^+(\tilde{\varepsilon}_F,k_y)\rangle+c_-|\phi^-(\tilde{\varepsilon}_F,k_y)\rangle+\sum_jc'_j|\phi_j^{(0)}(\tilde{\varepsilon}_F,k_y)\rangle,\label{eq_boundary}
\end{align}
with the decaying waves with the coefficient $c_\pm$ in the right side, where $\varepsilon_F$ and $\tilde{\varepsilon}_F$ indicate the Fermi level in the semiconducting and superconducting regions, respectively.
Here $k_y$ is restricted in the region where conducting channels are present in the metallic region.
Thus $p_x$ is much larger than $k_y$ because the Fermi pocket in the metallic monolayer is much smaller than that in the normal phase of the superconductor, $\varepsilon_F\ll\tilde{\varepsilon}_F$.
When the boundary condition is satisfied, the conservation of probability current density is fulfilled naturally because the equation for the condition is obtained by operating the velocity operator $v\sigma_x$ on both sides of Eq\ (\ref{eq_boundary}).
Thus we use the continuous condition of momentum $\hat{p}_x=i\hbar(\partial/\partial x)$ for calculating the coefficients,
\begin{align}
&-q_x|\psi_e^+(\varepsilon_F,k_y)\rangle+q_xr^{ee}|\psi_e^{-}(\varepsilon_F,k_y)\rangle-q_xr^{he}|\psi^{-}_h(\varepsilon_F,k_y)\rangle\nonumber\\
=&(-p_x+i\frac{\Delta}{v})c_+|\phi^+(\tilde{\varepsilon}_F,k_y)\rangle+(p_x+i\frac{\Delta}{v})c_-|\phi^-(\tilde{\varepsilon}_F,k_y)\rangle\nonumber\\
&+\sum_jc'_j\left(i\frac{\partial}{\partial x}\right)|\phi_j^{(0)}(\tilde{\varepsilon}_F,k_y)\rangle.\label{eq_continuous}
\end{align}
%
%
%
%
We eliminate $|\phi^{(0)}_j\rangle$ from the equations by taking the inner product with $|\phi_\pm\rangle$ due to the orthogonality and obtain four linear equations involving four coefficients; $r^{ee}$, $r^{he}$, $c_+$, and $c_-$.
These four equations enable us to acquire the analytic representation of Andreev reflection coefficient,
\begin{align}
r^{he}
=i\left(1-\left|\frac{\langle\varphi_e^+|\psi^{-}_e\rangle}{\langle\varphi_e^+|\psi_e^{+}\rangle}\right|^2\right)/\left(1+\left|\frac{\langle\varphi_e^+|\psi^{-}_e\rangle}{\langle\varphi_e^+|\psi_e^{+}\rangle}\right|^2\right),
\end{align}
with
\begin{align}
\frac{\langle\varphi_e^+|\psi^{-}_e\rangle}{\langle\varphi_e^+|\psi_e^{+}\rangle}=\frac{1-re^{-i\tau(\theta_p+\theta_q)}}{1+re^{-i\tau(\theta_p-\theta_q)}},\label{eq_factor}
\end{align}
where we use a relation of inner product $\langle\phi^\pm|\psi_e^+\rangle=\langle\phi^\mp|\psi_e^-\rangle^\ast=\pm i\langle\phi^\pm|\psi_h^-\rangle$ and the condition about wave number $\Delta/v,\;q_x\ll p_x$.

This analytic result shows that the Andreev reflection is suppressed if the Fermi level $\varepsilon_F$ gets close to the valence band top, i.e., $k_F\rightarrow0$, in the metallic region.
In this case, the Fermi surface and the pseudo-spin texture is represented by $S_F^{(1)}$ in Fig.\ \ref{fig_pseudo-spin}.
The pseudo-spin is polarized in the $z$ direction and orthogonal to that in the superconducting region where the pseudo-spin is parallel to the $xy$ plane.
Thus, $\langle\varphi_e^+|\psi^{\pm}_e\rangle$ is nearly the same amplitude for the incident and reflected waves in Eq.\ (\ref{eq_factor}).
The asymptotic form in the limit is given by
\begin{align}
r^{he}\sim\frac{vk_F\cos\theta_q}{m}.\label{eq_asymptotic_r}
\end{align} 
Therefore, the Andreev reflection probability $|r^{he}|^2$ increases as a function of $k_F^2$.

We also calculate the transmission probability in the junction with the normal phase. 
In this case, the reflected hole wave is absent and an electron wave with a positive velocity in the $x$ axis appears as a transmitted wave with the amplitude $\tilde{t}^{ee}$ in the scattering problem.
The stationary state satisfies the boundary condition,
\begin{align}
&|\psi_e^+(\varepsilon_F,k_y)\rangle+\tilde{r}^{ee}|\psi_e^-(\varepsilon_F,k_y)\rangle\nonumber\\
=&\tilde{t}^{ee}|\varphi^{+}_{e}(\tilde{\varepsilon}_F,k_y)\rangle,
\end{align}
and the continuous condition of momentum,
\begin{align}
&-q_x|\psi_e^+(\varepsilon_F,k_y)\rangle+q_x\tilde{r}^{ee}|\psi_e^-(\varepsilon_F,k_y)\rangle\nonumber\\
=&-p_x\tilde{t}^{ee}|\varphi^{+}_{e}(\tilde{\varepsilon}_F,k_y)\rangle,
\end{align}
where $|\varphi^{+}_{e}(\tilde{\varepsilon}_F,k_y)\rangle$ is the transmitted wave traveling to the positive $x$-axis.
Here, the reflection coefficient $\tilde{r}^{ee}$ is different from $r^{ee}$ in the metal-superconductor heterojunction.
Two linear equations are obtained by taking an inner product with the transmitted wave function and they enable us to calculate $\tilde{t}^{ee}$.
However, $\tilde{t}^{ee}$ is the transmission coefficient for a single channel at $k_y$.
Then, we correct the coefficient to obtain the transmission probability per energy by multiplying the ratio of the channel density in the superconductor to that in the metal.
The ratio can be calculated by that of velocity as $v/(v^2q_x/m)$, where these velocities can be obtained from Eq.\ (\ref{eq_Weyl_Hamiltonian}) and (\ref{eq_massive_Dirac}).
The transmission coefficient is given by
\begin{align}
t^{ee}=\sqrt{\frac{m}{vq_x}}\frac{2q_x}{p_x+q_x}\langle\varphi^{+}_{e}|\psi_e^+\rangle,
\end{align}
with
\begin{align}
\langle\varphi^{+}_{e}|\psi_e^+\rangle=\frac{1+re^{-i\tau(\theta_p-\theta_q)}}{\sqrt{1+r^2}}.
\end{align}
This procedure is not necessary for $r^{he}$ because the channel density is unchanged between the incident and reflected waves.
This analytic result indicates that the transmission probability increases with the matching of  pseudo-spin between the incident and transmitted waves.
The asymptotic form in the limit of $k_F\rightarrow0$ is given by
\begin{align}
t^{ee}\sim\frac{1}{p_x}\sqrt{\frac{mk_F\cos\theta_q}{v}},\label{eq_asymptotic_t}
\end{align}
with $q_x=k_F\cos\theta_q$ and $0<q_x$.

We briefly summarize the analytic results.
The Andreev reflection coefficient is proportional to the Fermi wave number $k_F$ in Eq.\ (\ref{eq_asymptotic_r}).
This dependence is attributed to the pseudo-spin texture of electrons forming Cooper pairs.
Since the pseudo-spin in the metallic TMDC reaches the fully polarization in the $z$-axis under $k_F\rightarrow0$, the pseudo-spin projection provides a unity as a factor in Eq.\ (\ref{eq_factor}).
Thus the Andreev reflection probability is zero at $k_F=0$ and increases as a linear function of $k_F$ due to the change of pseudo-spin projection.
In what follows, we numerically calculate the differential conductance in the lateral heterojunction of superconducting and semiconducting TMDC monolayers; NbSe$_2$ and MoSe$_2$, respectively.
The differential conductance is associated with the Andreev reflection.
We discuss the effect of pseudo-spin dependence to the differential conductance and compare it with the numerical result of conductance in the normal phase in the next section.

\section{Numerical calculation}\label{Sec_numerical}
\subsection{First-principles band structures}
\begin{figure}[htbp]
\begin{center}
 \includegraphics[width=80mm]{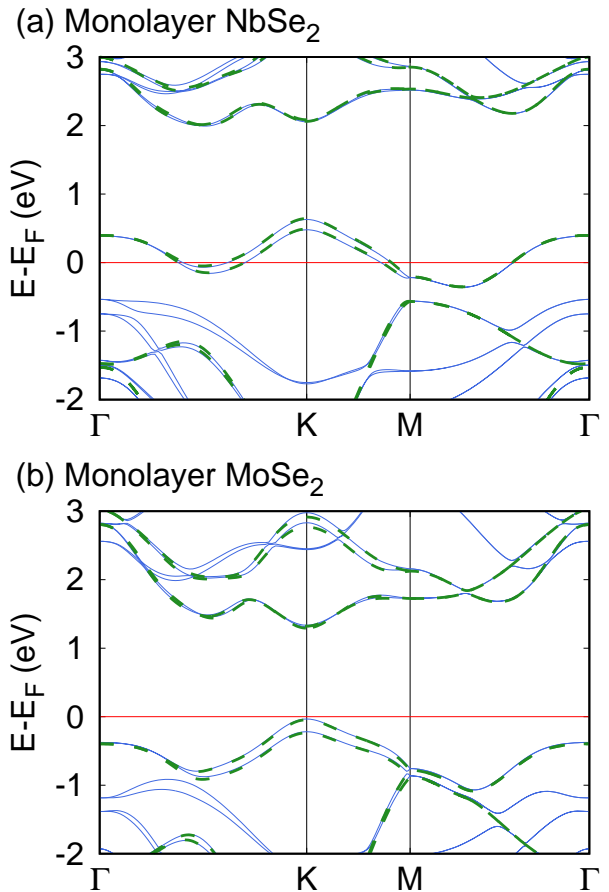}
\caption{
The band structure of (a)NbSe$_2$ and (b)MoSe$_2$.
The horizontal line indicates the Fermi energy of the pristine TMDC, and the dashed line is the band structure calculated by using a tight-binding model.
 }\label{fig_band}
\end{center}
\end{figure}

We investigate the electronic structure of MoSe$_2$ and NbSe$_2$ in 2H-type crystal structure by using a first-principles calculation based on density functional theory (DFT).
The band dispersion of MoSe$_2$ and NbSe$_2$ monolayers is calculated by using quantum-ESPRESSO,\cite{quantum-espresso} a first-principles calculation code, and shown in Fig.\ \ref{fig_band}.
Here, we adopt the lattice constants computed by using the lattice relaxation code in quantum-ESPRESSO.
The lattice parameters are $a_{\mathrm{Nb-Nb}}=3.476$\AA\ and $d_{\mathrm{Se-Se}}=3.514$\AA\ in NbSe$_2$, and $a_{\mathrm{Mo-Mo}}=3.319$\AA\ and $d_{\mathrm{Se-Se}}=3.343$\AA\ in MoSe$_2$, where $a_{M-M}$ ($d_{\mathrm{X-X}}$) is the horizontal (vertical) distance between nearest neighbor transition-metal (chalcogen) atoms.\cite{Habe2019-1}
We apply a projector augmented wave (PAW) method to the band calculation with a generalized gradient approximation (GGA) functional including spin-orbit coupling (SOC), and adopt the cut-off energy of plane wave basis 50 Ry, and the convergence criterion 10$^{-8}$ Ry.
The electronic bands are classified into two groups due to the eigenvalue of mirror operation because 2H-TMDC monolayers preserve mirror reflection symmetry in the out-of-plane axis.
In each band, electronic states are eigenstates of the mirror operator and characterized by the eigenvalue $\xi_z$.
Since the bands around the Fermi energy have $\xi_z=1$, we consider the electronic states with $\xi_z=1$ in what follows.

\begin{figure*}[htbp]
\begin{center}
 \includegraphics[width=170mm]{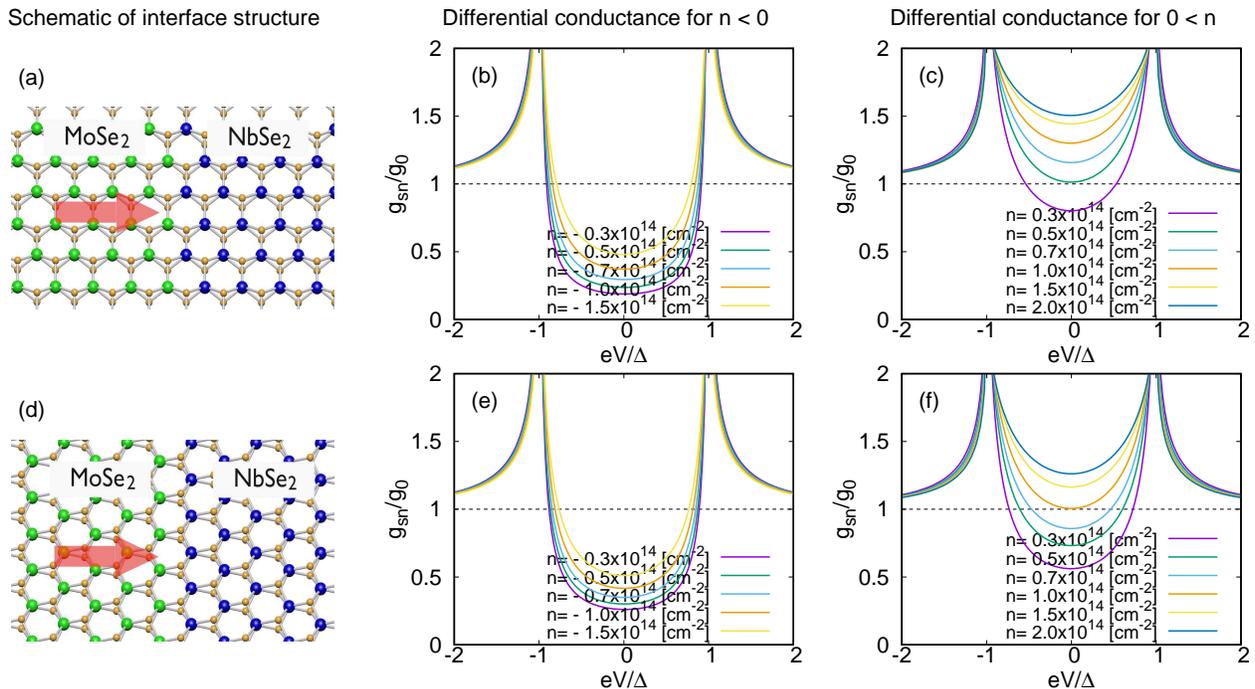}
\caption{
The differential conductance in the MoSe$_2$-NbSe$_2$ lateral heterojunction with the armchair interface in (b) and (c), and the zig-zag interface in (e) and (f). 
The schematics of interface structure in (a) and (b).
The arrow indicates the current direction.
 }\label{fig_diff_con}
\end{center}
\end{figure*}
We consider a tight-binding model to describe the local dynamics of electrons in the MoSe$_2$ (NbSe$_2$) monolayer where the Wannier orbitals are $p$-orbitals in Se and $d$-orbitals in Mo (Nb).
There are six Wannier orbitals with $\xi_z=1$ in the primitive unit cell; $|d_{3z^2-r^2}\rangle$, $|d_{xy}\rangle$, and $|d_{x^2-y^2}\rangle$ in Mo or Nb, and $|p^{+}_x\rangle$, $|p^{+}_y\rangle$, and $|p^{-}_z\rangle$ in Se.
Here $|p^{\pm}_\mu\rangle=(|p^{t}_\mu,\rangle\pm|p^{b}_\mu\rangle)/\sqrt{2}$ is the superposition of $p$-orbitals in the top Se and the bottom Se, $|p^{t}_\mu\rangle$ and $|p^{b}_\mu\rangle$, respectively.
The hopping integrals between these orbitals and on-site potential are computed from the first-principles bands in Fig.\ \ref{fig_band} by using Wannier90\cite{Wannier90}.
Since the DFT calculation underestimates the band gap in semiconductors,\cite{Qiu2013,Sejoong2017,Druppel2018} we use the charge density as a parameter instead of the Fermi energy.
The charge density can be controlled by using gates in experiments.\cite{Zhang2012-2} 
Each band splits into two due to the large spin-orbit coupling because of inversion symmetry-breaking in the crystal structure.
The effect of spin-orbit coupling is included as spin-dependent hopping integrals in the tight-binding Hamiltonian.
The spin-orbit coupling acts as a $k$-dependent Zeeman field in the $z$ axis due to mirror reflection symmetry.
Thus, electronic states split into two spin states, where the spin polarization direction is parallel to the $z$ axis.
The tight-binding model reproduces the first-principles bands as shown in Fig.\ \ref{fig_band}.
Here, the dashed lines are bands calculated by using the tight-binding model.

\subsection{Bogoliubov-de Gennes Hamiltonian}
Electronic states in the junction are described by a tight-binding model consisting of that of pristine MoSe$_2$ and NbSe$_2$ monolayers, where we consider two types of junctions with the zig-zag interface and the armchair interface as shown in Fig.\ \ref{fig_diff_con} (a) and (c).
The hopping matrix through the interface is assumed to be that for MoSe$_2$ because the difference of computed hopping matrix is smaller than 5 meV between MoSe$_2$ and NbSe$_2$. 
In this paper, we consider non-zero charge density induced by a homogeneous gate.
The Fermi energy aligns to that of the pristine monolayers with the distance far from the interface where it is calculated from
\begin{align}
n=\int\frac{d^2\boldsymbol{k}}{(2\pi)^2}\sum_{\alpha}\theta(E_F-E_{\alpha,\boldsymbol{k}}),
\end{align}
where $E_{\alpha,\boldsymbol{k}}$ is the energy dispersion of band $\alpha$ in Fig.\ \ref{fig_band}.
The bands in both of the monolayers are aligned for the Fermi energy to be matched.
In general, the potential fluctuation emerges near the interface of two monolayers and can be a contact resistance but we omit the effect for simplicity.
We also assume the atomically commensurate interface\cite{Li2015,Gong2014}  because a previous study show that a flat and commensurate heterojunction has a local minimum of free-energy even in the presence of mismatch in the lattice constant.\cite{Xie2018}
This allows us to use one-dimensional tight-binding Hamiltonian with a wave number parallel to the interface under the periodic boundary condition.
We represent the wave number by $k_y=2\pi m/L$ with an integer $m$, where $L$ is the perimeter along the $y$ axis.

The tight-binding Hamiltonian for each $k_y$ is obtained by using Fourier transformation in the $y$ axis,
\begin{align}
H^{(0)}_{k_y}=\sum_{j,\delta}&\left\{\hat{c}^\dagger_{k_y,j} {h}^{(0)}_{k_y,j}\hat{c}_{k_y,j}\right.\nonumber\\
&+\left.\left(\hat{c}^\dagger_{k_y,j+\delta} {t}^{(0)}_{k_y,j}\hat{c}_{k_y,j}+\mathrm{h.c.}\right)\right\},
\end{align}
where the on-site potential ${h}^{(0)}$ and the hopping matrix ${t}^{(0)}$ are $48\times48$ matrix, and $\hat{c}_{k_y,j}=(\hat{c}_{\uparrow,k_y,j},\hat{c}_{\downarrow,k_y,j})$ is a vector of the annihilation operators with $k_y$ on the site $j$ where the basis is defined by all the forty eight orbitals including the spin degree of freedom in the unit cell as shown in Fig.\ \ref{fig_schematic}.
Here, we adopt the hopping integrals computed from the first-principles bands of MoSe$_2$ (NbSe$_2$) as the on-site potential and the hopping matrix for $j\leq0$ ($0<j$).

We consider the superconducting states in NbSe$_2$ by using Bogoliubov-de Gennes (BdG) theory.
In this formalism, quasi-particle states are described by the BdG Hamiltonian.
The on-site potential ${h}_{k_y,j}$ and hopping matrix ${t}_{k_y,j}$ are defined on the basis of electron and hole states in the same manner as Eq.\ (\ref{eq_eff_BdG}),
\begin{align}
\begin{aligned}
{h}_{k_y,n_x}=&\begin{pmatrix}
{h}^{(0)}_{k_y,n_x}-\varepsilon_F&\Delta {I}\\
\Delta {I}&(is_y)(\varepsilon_F-{h}^{(0)}_{-k_y,n_x})^\ast(-is_y)
\end{pmatrix},\\
{t}_{k_y,n_x}=&\begin{pmatrix}
{t}^{(0)}_{k_y,n_x}&0\\
0&(is_y)(-{t}^{(0)}_{-k_y,n_x})^\ast(-is_y)
\end{pmatrix},\label{eq_tight-binding_model}
\end{aligned}
\end{align}
where $\Delta$ is the superconducting gap, ${I}$ is the identity matrix in the basis of $\hat{c}_{k_y,n_x}$, and $s_y$ is the Pauli matrix for the $y$ spin.
Here, the basis is the Nambu basis $\hat{b}=(\hat{c}_\uparrow,\hat{c}_\downarrow,\hat{c}^\dagger_\downarrow,-\hat{c}^\dagger_\uparrow)$ where the spin axis is chosen to be parallel to the $z$ axis.
The superconducting gap $\Delta$ can be estimated by $\Delta=(\pi/e^\gamma)T_c\simeq3.5T_c/2$ with the transition temperature $T_c$ and Euler's constant $\gamma$ according to Bardeen Cooper Schrieffer theory. 
The transition temperature is obtained as $T_c=3.0$ K for monolayer NbSe$_2$ experimentally.\cite{Lu2015,Wang2017}
In a bilayer NbSe$_2$, the transition temperature changes with the gate voltage\cite{Xi2017} but we omit a change of transition temperature by gating in this calculation.
Since the superconducting gap is absent in MoSe$_2$ region, we set $\Delta=0$ for $j\leq0$.

In the representation, incident electrons are reflected at the interface of the junction as far as the electron has an energy in the superconducting gap of NbSe$_2$ monolayer.
We briefly recall the discussion about the reflection process in the superconductor-metal junction in Sec.\ \ref{Sec_analyze_junction}.
There are two types of reflection processes classified by the charge of reflected particle.
The normal reflection means that the incident electron (hole) is reflected and goes back as an electron (hole).
The other is called Andreev reflection where the incident electron (hole) changes into a hole (electron) coming back.\cite{Blonder1982}
The Andreev reflection process is understood as that an incident electron transmits in the superconductor by forming a Cooper pair, the pair of electron bounded by an attractive force, with another electron which leaves a hole near the interface.
Thus, the charge transport property in the metal-superconductor junction is associated with the Andreev reflection.
The differential conductance $g_{sn}\equiv dI/dV$, which is defined by using the electric current $I$ and the source-drain bias voltage $V$, is given by 
\begin{align}
\frac{dI}{dV}=\frac{2e^2}{h} \sum_{l}(1-|r^{ee}_l|^2+|r^{he}_l|^2)
\end{align}
where $r^{ee}_l$ and $r^{eh}_l$ are the reflection coefficients for the normal reflection process and the Andreev reflection process of electrons in the conduction channel $l$.\cite{Blonder1982,Takane1991}

We study the electronic transmission between the MoSe$_2$ and NbSe$_2$ monolayers by using the two-terminal lattice Green's function method.
The heterojunction can be separated into two lead-regions and an interface region.
In the leads, the electronic states can be represented by $\hat{b}_{k_y,m}=\hat{b}_{k_y,0}e^{im\lambda}$, where $m$ indicates the position.
The phase factor $\lambda$ and the state vector $\hat{b}_{k_y,0}$ at an energy $E$ are described by
\begin{align}
\lambda\begin{pmatrix}
\hat{b}_{k_y,0}\\
\hat{b}_{k_y,-1}
\end{pmatrix}
=
\begin{pmatrix}
{t}_{k_y,n}^{-1}({h}_{k_y,n}-E)&-{t}_{k_y,n}^{-1}{t}^\dagger_{k_y,n}\\
1&0
\end{pmatrix}
\begin{pmatrix}
\hat{b}_{k_y,0}\\
\hat{b}_{k_y,-1}
\end{pmatrix}.
\end{align}
In the semiconductor, the eigenstates can be separated into those for electrons, $\hat{b}^e=(\hat{c}_\uparrow,\hat{c}_\downarrow,0,0)$, and holes, $\hat{b}^h=(0,0,\hat{c}^\dagger_\downarrow,-\hat{c}^\dagger_\uparrow)$, because of $\Delta=0$.
The reflection coefficients are computed from the lattice Green's function method\cite{Ando1991,Lewenkopf2013,Habe2015,Habe2016} in the multi-orbital tight-binding model in Eq.\ (\ref{eq_tight-binding_model}) for incident electron states.
The Andreev and normal reflection coefficients are distinguished by the amplitude of reflected waves, $\hat{b}^h$ and $\hat{b}^e$, respectively.

\begin{figure}[htbp]
\begin{center}
 \includegraphics[width=70mm]{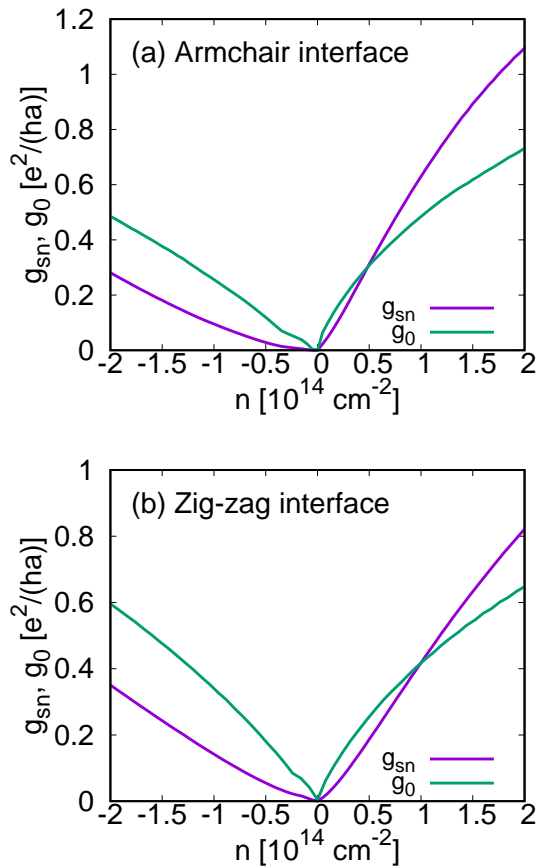}
\caption{
The charge density-dependence of the differential conductance and the normal conductance in the superconducting phase and the normal phase, respectively.
 }\label{fig_n-dependence}
\end{center}
\end{figure}
\subsection{Numerical results}
We show the differential conductance as a function of source-drain bias voltages $V$ in Fig.\ \ref{fig_diff_con}.
The vertical and horizontal axes indicate the normalized differential conductance $g_{sn}/g_0$ by the conductance $g_0$ in the normal phase of NbSe$_2$ and the source-drain voltage $eV$, respectively.
Here, the bias voltage is normalized by the gap energy $\Delta$, and $eV/\Delta=1$ implies that incident electrons have the energy matched to the edge of quasi-particle band. 
In the electron-doped hetelojunction $n<0$, the normalized differential conductance is suppressed and not sensitive to the charge density and the structure of interface, the armchair structure in (b) and the zig-zag structure in (e).
The differential conductance for $0<n$, on the other hand, strongly depends on the charge density, and it exceeds the normal conductance $g_0$ with increase in $n$.
This result indicates that the Andreev reflection is enhanced with increase in $n$.
The interface structure of heterojunction also quantitatively changes the differential conductance as shown in (c) and (f), and the armchair interface enhances it for hole-doped heterojunctions, $0<n$.
We also plot the $n$ dependence of the differential conductance in the superconducting phase and the normal conductance in the normal phase at $eV=0$ in Fig.\ \ref{fig_n-dependence}.
In the case of $n<0$, the normal conductance is always larger than the differential conductance  for both of the interface structures.

\begin{figure}[htbp]
\begin{center}
 \includegraphics[width=70mm]{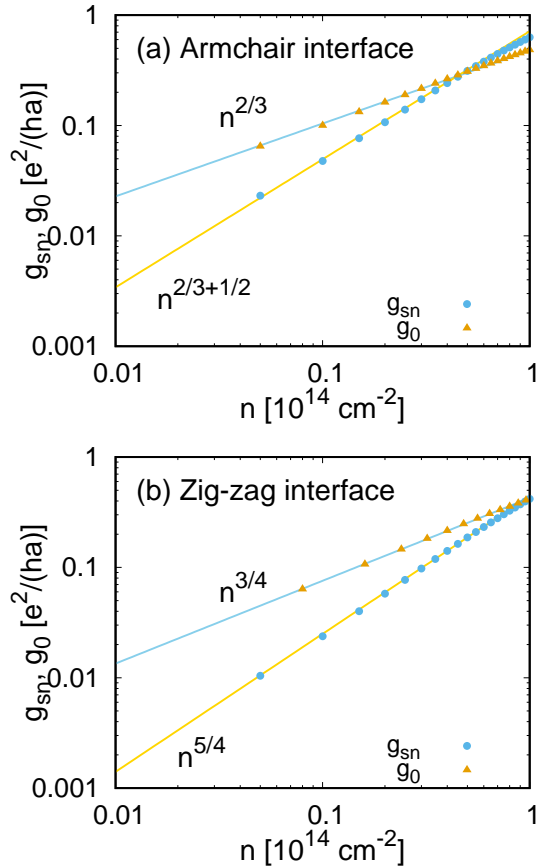}
\caption{
The log-log scale plot of $g_{sn}$ and $g_0$ with respect to the charge density in $0<n$. 
The solid lines are fitted to the numerical data (symbols) in the form of $n^\alpha$.
In the two cases, the difference of power indexes between $g_{sn}$ and $g_0$ is a half.
 }\label{fig_log-log}
\end{center}
\end{figure}
We also show the log-log scale plot of the $n$ dependence for the hole-doped heterojunction in Fig.\ \ref{fig_log-log}.
The numerical data align on a line in the form of $n^\alpha$ in both cases of $g_{sn}$ and $g_0$.
The gradient of lines is corresponding to the power index $\alpha$.
Although the index changes with the interface structure, the difference of the index between $g_{sn}$ and $g_0$ is a half regardless of the interface.
This difference causes the crossover from $g_{sn}<g_0$ to $g_{sn}>g_0$ with an increase of $n$ in Fig.\ \ref{fig_n-dependence}.

The analytic formula in Sec.\ \ref{Sec_analyze_junction} also provides the difference consistent with the numerical result.
The Andreev reflection probability $|r^{he}|^2$ and the transmission probability $|t^{ee}|^2$ are represented by functions of $k_F$ which is proportional to $\sqrt{n}$ because of the quadratic dispersion and the constant density of state.
Since $|r^{ee}|^2+|r^{he}|^2=1$ is fulfilled $eV$ inside the superconducting gap, the power index of $g_{sn}$ with respect to $n$ is that of $|r^{he}|^2$ plus a half.
Here the half is attributed to the integration with respect to $k_y$ for calculating $g_{sn}$ from $|r^{he}|^2$ at each $k_y$.
The transmission probability has the power index of $|t^{ee}|^2$ plus a half.
Therefore, it is confirmed that the difference of power indexes is a half by using the analytic results in Eq.\ (\ref{eq_asymptotic_r}) and (\ref{eq_asymptotic_t}) in Sec.\ \ref{Sec_analyze_junction}.

\section{Discussion}\label{Sec_discussion}
Finally, we consider the mismatch of power index between the numerical and analytic results.
It is attributed to the simplification of band structure of NbSe$_2$ in the analytic formulation.
The Fermi pocket in TMDCs is not isotropic around the K, K$'$, and $\Gamma$ points and warping due to three-fold rotation symmetry of the crystal structure.
The trigonal warping could change the wave vector dependence of Andreev reflection and transmission coefficients.
Moreover, the heterostructure with the Armchair interface allows the electronic transmission between the K or K$'$ valley in MoSe$_2$ and the $\Gamma$ valley in NbSe$_2$ as discussed in our previous paper.\cite{Habe2019-1}
This also can fluctuates the power index of $g_{sn}$ and $g_0$.
However, the difference of power index can remain robust even if the two types of interface are coexisting in the realistic heterojucntion.
Therefore, this crossover can be an experimental proof of the pseudo-spin triplet Cooper pair in the superconducting NbSe$_2$ monolayer.

\section{Conclusion}\label{Sec_conclusion}
In this paper, we have analyzed the pseudo-spin texture of electrons forming spin singlet Cooper pairs in superconducting TMDC monolayers and shown that the parallel pseudo-spin pair is realized in the superconducting monolayer.
The pseudo-spin polarizing direction varies with the direction and magnitude of the wave number.
Forming a heterojunction of superconducting and metallic monolayers, the pseudo-spin direction affects the electronic transmission and reflection probability.
The Andreev reflection, the electronic transmission with transforming a Cooper pair, vanishes in the limit of Fermi energy crossing the valence band edge.
We numerically calculated the conductance in the normal phase and the superconducting phase at each charge density.
We found that the conductance decreases with changing the phase from the normal one to the superconducting one in the low charge density region but it increases in the high charge density region.
This charge density dependence is attributed to the difference of pseudo-spin polarizing axis in the two TMDC monolayers, and it can be a proof of parallel pseudo-spin Cooper pair.
\bibliography{TMDC}
\end{document}